\documentclass[aps,twocolumn]{revtex4-1}

\usepackage{amsmath}
\usepackage{stmaryrd}
\usepackage{color}
\usepackage{mathrsfs}
\usepackage{rotating}
\usepackage{graphicx} 
\usepackage{amsfonts}
\usepackage{mathrsfs}
\usepackage[american]{babel}
\usepackage{pstricks}
\usepackage{pst-plot}
\usepackage{pst-grad}
\usepackage{pst-eps}
\usepackage{slashed}
\usepackage{pgfplots}
\usepackage{braket}
\usepackage{caption}
\usepackage{graphicx}
\usepackage{subfigure}

\newcommand{\ea}[1]{\begin{eqnarray}#1\end{eqnarray}}
\newcommand{\eq}[1]{\begin{equation}#1\end{equation}}

\begin{document}
\title{Vacuum polarization in asymptotically Lifshitz black holes}
\author{Gon\c{c}alo M. Quinta}
\email{goncalo.quinta@ist.utl.pt}
\affiliation{Centro Multidisciplinar de Astrof\'{\i}sica, CENTRA,
Departamento de F\'{\i}sica, Instituto Superior T\'ecnico - IST,
Universidade de Lisboa - UL, Avenida Rovisco Pais 1,
1049-001 Lisboa, Portugal\,\,\,\,}
\author{Antonino Flachi}
\email{flachi@phys-h.keio.ac.jp}
\affiliation{Department of Physics, and Research and Education Center for Natural Sciences, Keio University, Hiyoshi 4-1-1, Yokohama, Kanagawa 223-8521, Japan
\,\,\,}
\author{Jos\'{e} P. S. Lemos}
\email{joselemos@ist.utl.pt}
\affiliation{Centro Multidisciplinar de Astrof\'{\i}sica, CENTRA,
Departamento de F\'{\i}sica, Instituto Superior T\'ecnico - IST,
Universidade de Lisboa - UL, Avenida Rovisco Pais 1,
1049-001 Lisboa, Portugal\,\,\,}



\begin{abstract}
There has been considerable interest in applying the gauge/gravity duality to condensed matter theories with particular attention being devoted to gravity duals (Lifshitz spacetimes) of theories that exhibit anisotropic scaling. In this context, black hole solutions with Lifshitz asymptotics have also been constructed aiming at incorporating finite temperature effects. The goal here is to look at quantum polarization effects in these spacetimes, and to this aim, we develop a way to compute the coincidence limit of the Green's function for massive, non-minimally coupled scalar fields, adapting to the present situation the analysis developed for the case of asymptotically anti de Sitter black holes. The basics are similar to previous calculations, however in the Lifshitz case one needs to extend previous results to include a more general form for the metric and dependence on the dynamical exponent. All formulae are shown to reduce to the AdS case studied before once the value of the dynamical exponent is set to unity and the metric functions are accordingly chosen. The analytical results we present are general and can be applied to a variety of cases, in fact, to all spherically symmetric Lifshitz black hole solutions. We also implement the numerical analysis choosing some known Lifshitz black hole solutions as illustration.
\end{abstract}
\maketitle
\vspace{2mm}


\section{Introduction}
Ramifications of the AdS/CFT correspondence have been used to try to extend to non-relativistic condensed matter systems the idea that a strongly coupled theory can be described in terms of a gravitational weakly coupled dual (see Refs.~\cite{Hartnoll:2009sz,Herzog:2009xv} for reviews). 
One such non-relativistic incarnation of the correspondence has focused on the possibility of constructing a dual description of a class of critical phenomena in which phase transitions are governed by fixed points with dynamical scaling \cite{Kachru:2008yh,Taylor:2008tg}, 
\eq{
t\rightarrow \lambda^{z}t,~~~{\bf x} \rightarrow \lambda {\bf x},~~ z\neq 1,~
\label{scaling}
}
where $z$ is the dynamical exponent. The prototypical field theoretical example exhibiting the above anisotropic scaling is the Lifshitz field theory
\eq{
S= \int d^3x \left[
\ddot{\phi} -\kappa \left(\nabla^2\phi\right)^2
\right]~
\label{action}
}
that is invariant under (\ref{scaling}) with $z=2$.  

From the point of view of the duality, the question is then to find a general class of spacetimes that are dual to theories with non-trivial exponents $z \neq 1$. This problem has been addressed in Ref.~\cite{Kachru:2008yh}, where 
the following gravity solutions 
\eq{
ds^2 = \ell^2 \left(-r^{2z} dt^2 + r^2 d{\bf x}^2 + r^{-2}d^2r\right),
\label{geometry}
}
have been singled out. In the above coordinate system $0< r < \infty$, $d{\bf x}^2 = dx_1^2 + \cdots + dx_d^2$  and $\ell$ is the curvature scale of the geometry. For $z=1$ the metric reduces to $\mbox{AdS}_{d+2}$. Ref.~\cite{Kachru:2008yh} focuses on the case $z=2$ and $d=2$, appropriate to describe gravitational duals of $2+1$ dimensional field theories, like (\ref{action}). 
The geometry (\ref{geometry}) is nonsingular, and is invariant under the scale transformation
\eq{
t\rightarrow \lambda^{z}t,~~~r\rightarrow {r\over \lambda},~~~{\bf x} \rightarrow \lambda {\bf x},~~ z\neq 1~.
\label{scaling2}
}
Spacetimes like (\ref{geometry}) can be obtained as a solution of Einstein gravity augmented by a negative cosmological constant plus two- and three-form fields. Some details have been studied in Ref.~\cite{Kachru:2008yh}, where the two-point function and the holographic renormalization group flow have been worked out. 

Follow-up work, aiming at incorporating effects of finite temperature, has concentrated in finding Lifshitz black hole solutions, {\it i.e.} black holes asymptoting to the Lifshitz spacetime (\ref{geometry}). The general Euclidean form for a Lifshitz black hole can be expressed as
\eq{
ds^2 = \ell^2 \left(r^{2z} {f}(r) d\tau^2 + r^{-2} {u}(r) d^2r + r^2 d^2\Omega_2 \right),~
\label{lbh}
}
with $\tau = i t$ and which has the correct Lifshitz asymptotics (\ref{geometry}) if $f\rightarrow 1$ and $u\rightarrow 1$ as $r \rightarrow \infty$. The coordinate $\tau$ will also be regarded as periodic with period $\beta = 1/T_{\textrm{bh}}$ with $T_{\textrm{bh}}$ being the black hole temperature. Explicit solutions have been constructed in a variety of models mostly using numerical approximations, with some examples obtained analytically (see, for example, Refs.~\cite{Taylor:2008tg,Pang:2009ad,Danielsson:2009gi,Mann:2009yx,Bertoldi:2009vn,Balasubramanian,Brynjolfsson:2009ct,Pang:2009pd}). Black hole solutions with Lifshitz asymptotics were constructed in Ref.~\cite{Taylor:2008tg} and later generalized in Ref.~\cite{Pang:2009ad}. Ref.~\cite{Danielsson:2009gi} obtained black hole solutions for $z=2$ and $d=2$ by means of numerical approximation in the same model field theory proposed in \cite{Kachru:2008yh} (Einstein gravity with a cosmological constant plus two- and three-form gauge fields). A class of Lifshitz topological black holes were obtained in \cite{Mann:2009yx} also for $z=2$ and $d=2$. Lifshitz black holes with arbitrary $z$ were obtained in Refs.~\cite{Bertoldi:2009vn}. Other analytical solutions were found in Ref.~\cite{Balasubramanian}. Charged solutions with arbitrary $z$ have been obtained in Ref.~\cite{Brynjolfsson:2009ct} and generalizations obtained in Ref.~\cite{Pang:2009pd}. Solutions with $z=3/2$ have been obtained in Ref.~\cite{Cai:2009ac}. All the above cited solutions have the structure of Eq.~(\ref{lbh}). 

In this work we will take the general solution (\ref{lbh}) as a starting point and we will develop a method to obtain the coincidence limit of the Green function (the quantum vacuum polarization, $\langle \phi^2 \rangle$) for a bulk scalar $\phi$ on the background (\ref{lbh}). Such quantity is often taken as a starting point for the evaluation of the full quantum energy-momentum tensor and it is a useful preliminary step to investigate quantum fluctuations. Calculating the quantum vacuum polarization is usually technically complex and much work has been done to develop approximation schemes and numerical methods. 
Christensen \cite{christensen}  was the first to deal with this problem 
by calculating the vacuum expectation value of the
stress-energy tensor in a curved background
through the covariant point-splitting method.
Further developments were carried 
by Candelas and Howard \cite{Candelas:1980zt,Candelas:1984pg}  and later on by Anderson and collaborators \cite{Anderson:1989vg,Anderson2}. Some more recent calculations of $\braket{\phi^2}$ in different geometries can be found in \cite{EY, Breen, Tomimatsu, Cvetic}, for example.  In the present case, the calculation of $\langle \phi^2 \rangle$ should be adapted to the Lifshitz asymptotic structure. The case mostly relevant for us is that of asymptotically anti de Sitter solutions, for which the quantum vacuum polarization has been calculated in Ref.~\cite{flachi} and that we will follow closely here.

The starting point of our calculation is the known formula for the Green's function expressed as an infinite sum over the mode functions. The difficulties in evaluating its coincidence limit arise mainly due to the diverging behavior that prevent a direct numerical evaluation of the sums. A second problem arises due to the fact that exact solutions to the wave equation can only be found in some special cases (see Ref.~\cite{Cvetic:2015cca} for a recent example studying rotating black holes with subtracted geometry for which the vacuum polarization can be calculated analytically). The impossibility of finding explicit solutions translates into a difficulty of regularizing the coincidence limit and, in turn, into the impossibility of carrying out a direct numerical evaluation. A natural and widely used alternative is to approximate the solutions to the wave equation and use the approximate solutions to extract the divergences. This can be done by adopting some approximation scheme, like the WKB approximation, that proves to be sufficient to do the job. We should remark that while any approximation scheme for the solutions provides per se a way to approximate the vacuum polarization (or the energy-momentum tensor), a slow (or lack of) convergence can influence the result. In our case, the evaluation will be carried out exactly and the use of the WKB approximation serves only as intermediate step to extract the divergences. In the following sections we will detail the computations and report the intermediate steps that may ease the comparison with the asymptotically AdS case and serve as a check. While the method is valid for any black hole of the form (\ref{lbh}) with Lifshitz asymptotics, the numerical computation will be performed for some known solutions, thus specifying the form for the functions $f$ and $u$ and the value of the dynamical exponent $z$.


\section{Green's function in a Lifshitz spacetime}

Moving on to the problem at hand, let us consider a scalar field propagating on the Lifshitz black hole geometry (\ref{lbh}). The relevant action is
\eq{
S = {1\over 2} \int d^4x \sqrt{g} \left[
g^{\mu \nu} \partial_\mu \phi \partial_\nu \phi + m^2 \phi^2 + \xi R \phi^2
\right],~
}
where we included both a mass term $m$ and a coupling to the curvature. In the following, we will limit ourselves to consider a probe scalar, therefore the underlying theory leading to the solution (\ref{lbh}) will be unimportant. The form of the metric functions $f$ and $u$ will not be specified (except for their asymptotic behavior) and the analytical results that we will present are valid in general. Also, the computations will be carried out for a general value of $z$, which we will only specify at the end in the numerical evaluations, used to illustrate the results for specific examples. The method presented here refines and generalizes a similar one developed to compute the same quantity for anti de Sitter black holes \cite{flachi} that corresponds to $z=1$ and $f=u^{-1}$ and which the more general results presented here should reproduce. 

The Euclidean Green's function satisfies 
\eq{
\left(\Box - m^2 -\xi R \right) G_E(x,x') = - {\delta(x,x')\over \sqrt{g}}~,
}
with $g$ being the determinant of the metric tensor. Spherical symmetry allows one to express the Green's function 
as
\eq{
G_E(x,x') = {1\over \beta} \sum_{n,l} {2l+1\over 4\pi} e^{i\omega_n (\tau -\tau')} P_l(\cos\gamma)G_{nl}(r,r')~,
\label{Gansatz}
}
where $\omega_n= {2\pi n\over \beta}$ are the Matsubara frequencies, and $\cos\gamma =\cos\theta\cos\theta' + \sin\theta\sin\theta' \cos(\varphi-\varphi')$. Using the relation $\delta(\tau,\tau')= {1\over \beta} \sum_n e^{i\omega_n (\tau - \tau')}$, it is easy to express the differential equation for the radial Green function:
\begin{widetext}
\eq{
\left[{d^2\over dr^2} 
+ \left({3+z \over r} + {f'\over 2f} - {u'\over 2u}\right) {d\over dr}
-{u \over r^2}\left({\omega_n^2 \over r^{2z}f} +{l(l+1)\over r^2} +m^2 +\xi R \right)
\right] G_{nl}(r,r') =-\sqrt{{u \over f}} {\delta(r'-r) \over r^{z+3}}.
\label{eqt}
}
\end{widetext}
We have set $\ell=1$, thus fixed our units according to this choice. The solutions to the above equation can be expressed in terms of those of the homogeneous equation denoted here by $\mathscr{S}_{ln}(r)$, where $n$ and $l$ are the radial and angular quantum numbers, respectively.
There are two regions which are particularly important: near the horizon, and at infinity. In the latter case, the geometry has to recover the Lifshitz solution so $f\rightarrow 1$ and $u \rightarrow 1$. Therefore, in the asymptotic region, the homogeneous equation becomes
\eq{ \label{homsol}
\left[{d^2\over dr^2} 
+ {3+z\over r} {d\over dr} -\left({m^2+\xi R_{\infty} \over r^2} \right)
\right] \mathscr{S}^{\infty}_{nl}(r) =0~.
}
where $R_{\infty}= -2(z+1)^2 - 4$ is the Ricci scalar at large distance. The solution asymptotically is then
\eq{\label{homsol2}
\mathscr{S}^{\infty}_{nl}(r) \sim r^{\Delta_\pm}
}
where
\eq{
{\Delta_\pm}=
-{z+2\over 2} \pm
\sqrt{ \left({z+2 \over 2}\right)^2+m^2+\xi R_{\infty} }
}
Some care should be paid in selecting the correct solutions, or, in other words, the correct range of parameters. 
In the present case, one can easily observe that the parameters $m$, $\xi$ and $z$ (we assume $z \geq 1$ and $\xi \geq 0$) must satisfy the relation
\eq{
m^2 \geq -\left(
{z+2\over 2}
\right)^2 + 2 \xi \left((z+1)^2+2\right) \equiv \mu_\star^2,
}
and, in order for the solution asymptoting for $r^{\Delta_-}$ to fall off sufficiently rapidly, the condition 
\eq{
m^2 \leq 2 \xi \left((z+1)^2+2\right) = \mu_\star^2 + \left({z+2 \over 2}\right)^2
}
should also be satisfied. Within the region $\mu_\star^2 \leq m^2 \leq \mu_\star^2 + ((z+2)/2)^2$ both solutions are acceptable, while for $m^2 \geq \mu_\star^2 + ((z+2)/2)^2$ only the 
$\Delta_-$ solution falls off sufficiently rapidly. Setting $z=2$ and $\xi=0$ recovers known bounds. In the following, we will assume that the parameters satisfy the second inequality and take the solution relative to $\Delta_-$ as the only one acceptable.  

In the near horizon limit, we expect that $g_{\tau \tau} = r^{2z} f \to 0$ and $g_{rr} = u/r^2 \to \infty$ while at the same time $g_{\tau\tau}g_{rr} = r^{2z-2}f u \sim 1$. One way to find a solution in this limit is to rescale the coordinates as
\eq{
dr_* = dr/f^*
}
with
\eq{
f^*=r^{z+1}\sqrt{{f \over u}}~,
}
which allows us to rewrite the homogeneous equation in the form
\ea{
&&\bigg[ {d^2 \over dr_*^2}  - (r^{2z}f)\bigg( {l(l+1) \over r^2} + f^{*'} {r^{z+1} \over \sqrt{f\,u}} +m^2+\xi R \bigg)\nonumber \\
&& \hspace{10mm} - \omega^2_n\bigg](r\mathscr{S}_{nl}(r)) = 0~,
}
where the derivative is with respect to the variable $r$. Thus the horizon limit implies that
\eq{
\bigg[ {d^2 \over dr_*^2} - \omega^2_n\bigg](r\mathscr{S}^{h}_{nl}(r)) = 0~,
}
leading to
\eq{
\mathscr{S}^{h}_{nl}(r) \sim {e^{\pm \omega_n r_*} \over r}~.
}
The case with the $+$ sign is the solution regular at the horizon.


Following the usual notation, the solutions will be indicated as $p_{nl}(r)$ and $q_{nl}(r)$ according to the regularity at the horizon and at infinity, respectively, in which case the radial Green function can be shown to have the form
\eq{
G_{nl}(r',r)={1 \over r^{z+3}} \sqrt{u\over f}~{p_{nl}(r_<)q_{nl}(r_>)\over q_{nl}p'_{nl}-p_{nl}q'_{nl}}
}
where $r_>$ and $r_<$ denote respectively the highest and lowest of the set $\{r',r\}$, and 
where the primes denote differentiation with respect to the variable $r$. The Green function can thus be expressed in the following form
\ea{
\label{G2}
G_E(x,x') & = & {1\over \beta} \sum_{l,n} {2l+1\over 4\pi} e^{i\omega_n (\tau -\tau')} P_l(\cos\gamma)\nonumber \\
&& \hspace{-10mm} \times {1 \over r^{z+3}}\sqrt{u\over f} {p_{nl}(r_<)q_{nl}(r_>)\over q_{ln}(r)p'_{ln}(r)-p_{ln}(r)q'_{ln}(r)}~.
}
As we have explained, due to the diverging behavior, it is not possible to evaluate numerically the above expression before regularization. To by-pass this problem we shall adopt the usual procedure of approximating the solutions by means of the WKB approach and cast (\ref{G2}) in a form suitable for renormalization. 
The WKB form of the solutions is
\eq{
\mathscr{S}_{ln}(r) = r^{a} W^{b} \exp\left({\pm \int_{r_s}^{r} W^{c}(u) \chi(u)} du\right)~,
\label{wkbansatz}
}
where $\chi(u)= u^{d} g^{e}(u) f^{o}(u)$. The $+$ and $-$ signs refer to the solutions regular the horizon and infinity, respectively. By inserting this ansatz in the homogeneous equation, we are left with the associated differential equation for the function $W$. We choose the powers of each term in Eq.~(\ref{wkbansatz}) in such a way so as to eliminate the terms with ($\pm$)-signature. Additionally, we require that the term $r^a$ be consistent with the limit of Eq.~(\ref{homsol2}). All of these restrictions amount to setting
\ea{
a&=&-(z+2)/2~,~~b=-1/2~,~~c=1~,\nonumber\\
d&=&-	1~,~~e=1/2~,~~o=-1/2~,\nonumber
}
and the homogeneous equation becomes
\eq{
W^2 =  \varpi + \sigma + a_1 {W'\over W} +a_2 {W^{'2}\over W^2} +a_3 {W^{''}\over W}~,
\label{wkbeqt}
}
where
\ea{
\varpi &=& \left[\left(l+{1\over 2}\right)^2-{1\over 4}\right] {f \over r^2}+{\omega_n^2\over r^{2z}}~,\nonumber\\
\sigma &=& \left(m^2 +\xi R\right)f +\left({z\over 2}+1\right)^2 {f\over u}
+ {r\over 2}\left(1+ {z \over 2}\right) {f'\over u} \nonumber \\
& & 
- {r \over 2}\left(1+{z \over 2}\right) 
{f u'\over u^2}~,\nonumber
}
and
\ea{
a_1 &=& {r\over 2} {f\over u} -{r^2\over 4} {f \over u} \left[{u'\over u} - {f'\over f}\right]~,\nonumber\\
a_2 &=& -{3r^2\over 4} {f\over u}~,\nonumber\\
a_3 &=& {r^2\over 2}{f\over u}~.\nonumber
}
One may notice that the ansatz (\ref{wkbansatz}) follows the correct asymptotic behavior and all the above formulae reduce to the those for AdS in the appropriate limit $z=1$ and $f=u^{-1}$. 

It is now possible to express the solution iteratively
\ea{
W=W^{(0)} + W^{(1)} + \cdots~.
}
Following \cite{flachi}, we will work at next-to-leading order in the WKB, and approximate the solutions as follows
\eq{\label{Wwkb}
{1\over \tilde{W}_n(l)} = {1\over \Phi(l)^{1/2}}- {\Psi(l)\over 4\Phi(l)^{3/2}}~,
}
with $\tilde{W}_n(l)$ representing the approximated value of $W_n(l)$ and
\ea{\label{Psi}
\Phi(l) &=& \varpi +\sigma~,\nonumber\\
\Psi(l) &=& a_1 {\Phi'\over \Phi} + \left( {a_2-a_3\over 2} \right) \left({\Phi'\over \Phi}\right)^2 +a_3 {\Phi^{''}\over \Phi}~,
}
where, for future convenience, we have adopted the notation $\tilde{W}_n(l)$ and $\Phi(l)$ which makes explicit the argument $l$ while carrying an implicit dependence on $r$. The derivatives are also taken with respect to the variable $r$.


\section{Regularization of the Coincidence limit}

In this section we will construct the regulated coincidence limit of the Green's function.
First of all, we take the partial coincidence limit $(r',\Omega') \to (r,\Omega)$, define $\alpha = 2\pi / \beta$ and insert the WKB ansatz Eq.~(\ref{wkbansatz}) in Eq.~(\ref{G2}) to obtain
\eq{\label{G}
G_E(x,x') = {\alpha\over 8 \pi^2} \sum^{\infty}_{n=-\infty} e^{i n \alpha \varepsilon} \sum^{\infty}_{l=0}{(l+1/2)\over r^{z+2} W_{n}(l)} ~,
}
where we have defined $\varepsilon = \tau - \tau'$. Both sums, over $l$ and $n$, are divergent. We first deal with the $l$-summation by noticing that, since the complex exponential is essentially $\delta(\varepsilon)$, the Green function is devoid of divergences so long as $\varepsilon \neq 0$, or equivalently $X \neq X'$. Thus we are free to add multiples of $\delta(\varepsilon)$, which as long as $\varepsilon\neq 0$ will still be 0 so the final result will not be altered. As in Ref.~\cite{flachi}, we may choose multiples of the form
\eq{
{\alpha \over 8\pi^2} \sum_n e^{i n \alpha \varepsilon} \sum_l R_l(r)
}
where $R_l(r)$ is arbitrary and independent of $n$. Subtracting this term in Eq.~(\ref{G}) and looking at the asymptotic behavior of $W_{n}(l)$ for large $l$
\eq{
W_n(l) \sim \tilde{W}_n(l) \sim \left(l + {1 \over 2}\right){\sqrt{f} \over r} + \mathcal{O}\left[\left(l + {1 \over 2}\right)^{-1}\right]
}
we see that it is enough to set $R_l(r)=1/(r^{z+1}\sqrt{f})$ in order to cancel the asymptotic terms and thus removing the divergence in the summation over $l$, resulting in
\eq{\label{Gregl}
G_E(x,x') = {\alpha\over 8 \pi^2} \sum^{\infty}_{n=-\infty} e^{i n \alpha \varepsilon} \sum^{\infty}_{l=0}\left[{l+1/2\over r^{z+2} W_{n}(l)}-{1 \over r^{z+1} \sqrt{f}}\right]~.
}
While the divergences due to the $l$ summation can be eliminated without the need for renormalization, the more serious divergences appear in the coincidence limit $\varepsilon \rightarrow 0$ due to the summation over $n$.
These UV divergences can be traced back to fact that the coincident limit of the Green's function comes from the product of quantum fields which are being evaluated at the same spacetime point. Such divergences can be cured by subtraction of appropriate counter-terms. The problem of isolating the divergent terms for a general spherical symmetric metric has been addressed in general in Ref.~\cite{christensen}, where general formulae have been obtained. In the present case, the divergent terms are given by the limit $\varepsilon=0$ of the following expression
\ea{\label{Gdiv1}
G_{E\,\rm div.}(x,x') & = & {1 \over 16\pi^2}\bigg\{ {4 \over \varepsilon^2 f r^{2z}} + \left(m^2+\left(\xi-{1\over 6}\right)R\right)\times \nonumber \\
& & \hspace{-15mm} \times 
\left(\ln\left[{m^2 f \varepsilon^2 r^{2z} \over 4}\right]+2\gamma\right) - m^2 + {r^2 \over 6u}{f'^2 \over f^2} -{r^2 \over 6u}{f'' \over f}
\nonumber \\
& & \hspace{-15mm}
 \nonumber \\
& & \hspace{-15mm} 
+ {z r \over 6u}{u' \over u}+{r^2 \over 12u}{f' \over f}{u' \over u} -{r \over 2u}{f' \over f} - {2z \over 3u}\bigg\}\,.
}
where $\gamma$ is Euler's constant. Some finite terms are also included, as is common practice, although for practical purposes they are irrelevant. 

In order to renormalize the coincidence limit of the Green function, we must subtract the above expression from Eq.~(\ref{Gregl}) in the limit $\varepsilon=0$. While formally finite, it is convenient to recast the counter-terms in a form more suitable for the subsequent evaluation by means of 
the Abel-Plana summation formula, as done, for instance, in \cite{Anderson:1989vg}, leading to
\ea{\label{Gdiv2}
\lim_{x'\to x} & & G_{E\,\rm div.}(x,x') = {\alpha \over 8 \pi^2}\bigg[\Delta_1+ \Delta_{2}+\Delta_{3}+\Delta_{4} \nonumber \\
& & 
-\sum^{\infty}_{n=1}\bigg({2\omega_n \over r^{2z}f} + \left[m^2-\left(\xi-{1\over 6}\right)R\right]{1\over \omega_n}\bigg)\bigg].~~~~~~
}
In the above expression we have defined
\begin{widetext}
\eq{
\Delta_1 \equiv -\sum^{\infty}_{n=1}\left[ {2\over r^{2z}f}\left(\sqrt{\omega^2_n+m^2r^{2z}f} -\omega_n-{m^2r^{2z}f \over 2\omega_n}\right) +\left(\xi-{1\over 6}\right)R\left({1\over \sqrt{\omega^2_n+m^2r^{2z}f}}-{1\over \omega_n}\right) \right]
\nonumber}
and 
\ea{
\Delta_{2} &\equiv& {m^2 \over 2\alpha} \ln\left(m^2r^{2z}f\right) - {m^2 \over \alpha} \ln\left(\alpha + (\alpha^2+m^2r^{2z}f)^{1/2}\right) \nonumber \\
&& \hspace{10mm}+{2i\over r^{2z}f} \int^{\infty}_{0}{dt \over e^{2\pi t}-1}\left(\left[(1+it)^2\alpha^2+m^2r^{2z}f\right]^{1/2}-\left[(1-it)^2\alpha^2+m^2r^{2z}f\right]^{1/2}\right),\nonumber}

\ea{
\Delta_{3} &\equiv& {\left(\xi-{1\over 6}\right)R} \bigg[{\ln\left(m^2 r^{2z}f\right)\over \alpha}-{2\over \alpha}\ln\left(\alpha + (\alpha^2+m^2r^{2z}f)^{1/2}\right)+{1\over \sqrt{\alpha^2+m^2r^{2z}f}} \nonumber\\
&& \hspace{10mm} + 2\alpha i \int^{\infty}_{0}{dt \over e^{2\pi t}-1}\left({1\over \left[ (1+it)^2\alpha^2+m^2r^{2z}f\right]^{1/2}}-{1\over \left[(1-it)^2\alpha^2+m^2r^{2z}f\right]^{1/2}}\right)\bigg],\nonumber}

\ea{
\Delta_{4} &\equiv& {1\over 2\alpha}\left({r^2 \over 6u}{f'^2 \over f^2} -{r^2 \over 6u}{f'' \over f}+{z r \over 6u}{u' \over u}+{r^2 \over 12u}{f' \over f}{u' \over u} -{r \over 2u}{f' \over f} - {2z \over 3u}-m^2\right).
\nonumber}

From the above results, the coincidence limit is readily obtained by subtracting the divergent expression (\ref{Gdiv2}) from the non-renormalized result (\ref{Gregl}). After subtraction, the coincidence limit can be safely taken leading to a regular (finite) expression for the coincidence limit
\ea{\label{phi}
\braket{\phi^2(x)} & = & G^{(\rm ren.)}_E(x,x)\nonumber \\
& = &{\alpha \over 8 \pi^2}\bigg\{ \sum^{\infty}_{l=0}\left({l+1/2\over r^{z+2} W_{0 }(l)}-{1 \over r^{z+1} \sqrt{f}}\right) + \sum^{\infty}_{n=1}\bigg[2\sum^{\infty}_{l=0}\left({l+1/2\over r^{z+2} W_{n}(l)}-{1 \over r^{z+1} \sqrt{f}}\right)
\nonumber \\
& & \hspace{10mm} + {2\omega_n \over r^{2z}f} + \left[m^2-\left(\xi-{1\over 6}\right)R\right]{1\over \omega_n}\bigg]-\Delta_1-\Delta_{2}-\Delta_{3}-\Delta_{4}\bigg\}.
}
\end{widetext}


\section{Regularity and Summations}

\subsection{General procedure}

Although Eq.~(\ref{phi}) is finite by construction, the renormalized vacuum polarization, as written in (\ref{phi}) is not yet suitable for straightforward numerical evaluation. First of all, individual pieces are divergent. Thus, it is more than instructive to see how these can be combined in order for the divergences to cancel. Secondly, the summing over the angular and radial eigenvalues is numerically nontrivial, and it can expedited by appropriately expressing the sums. In this section, we will prove the regularity and outline how the summations are performed.

A useful way to proceed is to add and subtract the WKB counterpart of $\langle \phi^2(x) \rangle$. While this manipulation is only formal, it is clear that the reminder of the WKB approximation, $\delta \langle \phi^2(x) \rangle = \langle \phi^2(x) \rangle - \langle \phi^2(x) \rangle_{\textrm{WKB}}$, is regular and can be calculated numerically, modulo a convenient combination of the individual terms as we will describe below. Following previous work \cite{Anderson:1989vg,flachi}, we write 
\ea{\label{phiterms}
\braket{\phi^2(x)} = {\alpha \over 8 \pi^2}\{\Upsilon_0+\Sigma_1+\Sigma_2-\Delta\}
}
where we have defined $\Delta = \Delta_1+\Delta_{2}+\Delta_{3}+\Delta_{4}$,
\ea{
\Sigma_1 & \equiv & \sum^{\infty}_{l=0}\left(l+{1\over 2}\right)\left({1\over r^{z+2} W_{0}(l)}-{1\over r^{z+2} \tilde{W}_{0}(l)}\right) \nonumber \\
& & + 2\sum^{\infty}_{n=1}\sum^{\infty}_{l=0}\left(l+{1\over 2}\right)\left({1\over r^{z+2} W_{n}(l)}-{1\over r^{z+2} \tilde{W}_{n}(l)}\right)\,, \nonumber \\
\Sigma_2 & = & \sum^{\infty}_{n=1}\left[2\Upsilon_n+{2\omega_n \over r^{2z}f} + \left[m^2-\left(\xi-{1\over 6}\right)R\right]{1\over \omega_n}\right] \label{Sigma2}
}
and
\eq{
\Upsilon_n \equiv \sum^{\infty}_{l=0}\left({l+1/2\over r^{z+2} \tilde{W}_{n}(l)}-{1 \over r^{z+1} \sqrt{f}}\right)~.
}
Every term except for $\Sigma_2$ is explicitly finite, so we need only to verify that all the divergences in this term cancel. Obvious divergences come from the second and third terms, proportional to $\omega_n$ and $\omega_n^{-1}$, in (\ref{Sigma2}) and should cancel with terms coming from $\Upsilon_n$. In order to isolate these diverging terms, we can apply once more the Abel-Plana formula and recast $\Upsilon_n$ as
\ea{\label{Ups}
\Upsilon_n & = & {1\over r^{z+2}}\bigg\{{1\over 4\tilde{W}_n(0)}+\left[\int^{\infty}_0\left({\tau+1/2\over \tilde{W}_{n}(\tau)}-{r\over \sqrt{f}}\right)d\tau - {r\over 2\sqrt{f}}\right] \nonumber \\
& & + i\int^{\infty}_0{d\tau \over e^{2\pi \tau}-1}\left({i\tau+1/2 \over \tilde{W}_n(i\tau)}-{-i\tau+1/2 \over \tilde{W}_n(-i\tau)}\right)\bigg\}~.
}

The first term in Eq.~(\ref{Ups}) can be expressed in terms of the following Epstein-Hurwitz $\zeta$-function
\eq{
\mathcal{Z}_q \equiv \sum^{\infty}_{n=1} \left(\omega^2_n+r^{2z}\sigma(r)\right)^{-q/2}
}
as some straightforward calculations show
\begin{widetext}
\ea{\label{P1}
{1\over 2r^{z+2}}\sum^{\infty}_{n=1}{1\over \tilde{W}_n(0)} 
& = & {1\over 4r^2} \bigg\{2\mathcal{Z}_1+zr^{2z-2}\left[a_1r-2z\left({a_2-a_3\over 2}\right)-(2z+1)a_3\right]\mathcal{Z}_3 \nonumber \\
& & 
- r^{4z-2}
\left[
\left({a_1r\over 2}-2\left({a_2-a_3\over 2}\right)z\right)\left(2z\sigma+r\sigma'\right) 
-a_3\left(z(2z+1)\sigma-{r^2\sigma''\over 2}\right)
\right]\mathcal{Z}_5 \nonumber \\
&&- \left({a_2-a_3\over 2}\right){r^{4z}\over 2}
\left(2z\sigma r^{z-1} +r^z\sigma'\right)^2\mathcal{Z}_7 \bigg\}.
}
\end{widetext}
Having expressed the result in terms of the above Epstein-Hurwitz $\zeta$-function, isolating the divergences is only a matter of simple power-counting. In Eq.~(\ref{P1}), only the term multiplied by $\mathcal{Z}_1$ is responsible for the divergences, that can be extracted by expanding $\mathcal{Z}_1$ in powers of $\omega_n$ and retaining the terms proportional to $c_1 \omega_n + c_2 \omega_n^{-1}$. Simple steps lead to the divergent piece of (\ref{P1}), denoted by ${\rm div}_1$
\eq{\label{divP1}
{\rm div}_1
={1 \over 2r^2 \omega_n}~.
}

The next contribution to $\Sigma_2$ is the term in square brackets in Eq.~(\ref{Ups}), namely
\ea{\label{P2}
{2\over r^{z+2}}\sum^{\infty}_{n=1}\left[\int^{\infty}_0\left({\tau+1/2\over \tilde{W}_{n}(\tau)}-{r\over \sqrt{f}}\right)d\tau - {r\over 2\sqrt{f}}\right]
& = & A_1 + A_2
}
that we have rearranged as the sum of two pieces
\ea{
A_1 
&\equiv& {2\over r^{z+2}}\sum^{\infty}_{n=1}\left[\int^{\infty}_0\left({\tau+1/2\over \Phi(\tau)^{1/2}}-{r\over \sqrt{f}}\right)d\tau - {r\over 2\sqrt{f}}\right] \nonumber ~, \\
A_2
&\equiv& -{1\over 2r^{z+2}}\sum^{\infty}_{n=1}\int^{\infty}_0 {(\tau+1/2) \, \Psi(\tau)\over \Phi(\tau)^{3/2}} \, d\tau \nonumber ~.
}
The first term above can be easily evaluated by direct integration leading to
\eq{
A_1
= -{2 \over r^{2z} f} \mathcal{Z}_{-1}~,
}
from which the divergent contribution ${\rm div}_2$ can be extracted:
\eq{\label{divP21}
{\rm div}_2
= {2 \over r^{2z}f}\left(\omega_n + {r^{2z} \sigma \over 2\omega_n}\right)
}
For the other term, $A_2$, inserting the explicit expression for $\Psi$, integrating over $\tau$, and summing over the frequencies gives
\begin{widetext}
\ea{
A_2
& = & {a_1 \over 6f} \bigg\{\left({4+2z \over r}-2{f' \over f}\right)\mathcal{Z}_1 -r^{2z-1}(2z\sigma +r \sigma')\mathcal{Z}_3\bigg\} 
\nonumber \\& & 
- {a_2 -a_3\over 60f}\bigg\{\left({4(3z^2+4z+8) \over r^2}-{8(z+4) \over r}{f' \over f} + 8{f'^2 \over f^2}\right)\mathcal{Z}_1 
\nonumber \\& & 
-4r^{2z-1}\left({(3z+2) \over r}-{f' \over f}\right)(2z\sigma +r \sigma')\mathcal{Z}_3 + 3 r^{4z-2}(2z\sigma +r \sigma'^2)\mathcal{Z}_5\bigg\} 
\nonumber \\& & 
- {a_3 \over 6f}\bigg\{\left({12+2z(2z+1) \over r^2}-{8 \over r}{f' \over f}+2{f'' \over f}\right)\mathcal{Z}_1 - r^{2z-2}(2z(2z+1)\sigma-r^2\sigma'')\mathcal{Z}_3\bigg\}~.
}
\end{widetext}
In this case to, the only divergent contribution ${\rm div}_3$, which simple steps allow us to isolate, comes from $\mathcal{Z}_1$:
\ea{\label{divP22}
{\rm div}_3
& = & {1\over 6\omega_n} \bigg\{{z(4-z) \over 2u}-{z\over 2}{r\over u}{f'\over f}-{r^2\over u}{f''\over f} +{r^2\over 2u}{f'^2\over f^2} \nonumber \\
& & 
+{r\over 2u}{u'\over u}\left(r{f'\over f}-(z+2)\right)\bigg\}~.
}
The last term in Eq.~(\ref{Ups}) to consider is
\eq{\label{P3}
A_3  \equiv 
{2i\over r^{z+2}}\sum^{\infty}_{n=1}\int^{\infty}_0{d\tau \over e^{2\pi \tau}-1}\left({i\tau+1/2 \over \tilde{W}_n(i\tau)}-{-i\tau+1/2 \over \tilde{W}_n(-i\tau)}\right)~.
}
Observing that the dominant contribution to the integral comes from the $\tau \sim 0$ region of integration, we make the expansion
\eq{\label{expP3}
\left({i\tau+1/2 \over \tilde{W}_n(i\tau)}-{-i\tau+1/2 \over \tilde{W}_n(-i\tau)}\right) = -i \sum^{\infty}_{j=1} c_{nj} \tau^{2j-1}
}
for small $\tau$ and proceed by direct integration. Some calculations give

\ea{\label{P3f}
A_3
& = & {1\over r^{z+2}}\sum^{\infty}_{n=1}\sum^{\infty}_{j=1} {(-1)^{j-1} \over 2j} c_{nj}  B_{2j}
}
where $B_{2j}$ are the Bernoulli numbers and the coefficients $c_{nj}$ are those coming from the Taylor expansion of the integrand in Eq.~(\ref{P3}), 
\eq{\label{Cnj}
c_{nj} \equiv {i^{2j} \over (j-1)!}{d^{j-1} \over dx^{j-1}}\left({2 \over \tilde{W}_n(x)} - {1 \over j}{\tilde{W}'_n(x) \over \tilde{W}^2_n(x)}\right)\bigg|_{x=0}.
}

Identification of the divergent part in Eq.~(\ref{P3f}) is possible by looking at the asymptotic behavior of the coefficients $c_{nj}$ for large $n$.
Using Eqs.~(\ref{Wwkb})-(\ref{Psi}) it is straightforward to see that for $n \gg 1$ only the $j=1$ term in Eq.~(\ref{Cnj}) leads to a divergence,
\eq{\label{Cn1}
c_{n1} = -{2r^z \over \omega_n}.
}
Then, simple steps give the divergent contribution ${\rm div}_4$ arising from (\ref{P3})
\eq{\label{divP3}
{\rm div}_4
= -{1 \over 6r^2\omega_n}~.
}
From here it is a trivial exercise to sum (\ref{divP1}), (\ref{divP21}), (\ref{divP22}) and (\ref{divP3}) and prove that the final expression is regular.

\subsection{Numerical computations}

Having demonstrated the regularity of the results, the left-over task is to compute the summations over $n$, a problem that we can approach exactly as done for the AdS case and that we repeat here for the reader's convenience. 
Practically the problem has been reduced to calculating the zeta functions $\mathcal{Z}_q$. Only $\mathcal{Z}_1$ and $\mathcal{Z}_{-1}$ contain divergences, and, using the proof of the regularity, we can simply regulate these functions by subtracting the corresponding diverging contributions. This translates to the following definition
\ea{
\tilde{\mathcal{Z}}_{-1} & = & \alpha \sum^{\infty}_{n=1}\left(\sqrt{n^2+v^2}-n-{v^2\over 2n}\right),~~~~~ \\
\tilde{\mathcal{Z}}_{1} & = & \alpha^{-1} \sum^{\infty}_{n=1}\left({1 \over \sqrt{n^2+v^2}}-{1 \over n}\right), \label{regz}
}
with $v^2 \equiv {r^{2z}\sigma \over \alpha^2}$. Numerical evaluation can then be performed in different ways, depending on the magnitude of $v^2$. For large $v^2$, one may adopt the Chowla-Selberg formula \cite{elizalde,cs}
to recast the zeta-functions as
\ea{
\mathcal{Z}_q & = & \alpha^{-q}\bigg(-{v^{-q}\over 2}+{\sqrt{\pi}\,\Gamma((q-1)/2)\over 2 \Gamma)(q/2)}v^{1-q} \nonumber \\
& & 
+ {2\pi^{q/2}\over \Gamma(q/2)}v^{(1-q)/2}\sum^{\infty}_{p=1}p^{(q-1)/2}K_{(q-1)/2}(2\pi p v)\bigg)\,,\nonumber
}
which is regular for any $q \neq -1, 1$. For either $q=-1, 1$, following the logic we have explained before, one may simply subtract the divergent portion obtaining
\ea{
\tilde{\mathcal{Z}}_{-1} & = & \alpha \bigg({1\over 12} - {v\over 2} +{v^2\over 4}\left[1-2\ln(v/2)+\gamma\right] \nonumber \\
&& -{v\over \pi}\sum^{\infty}_{p=1}{K_{-1}(2\pi p v)\over p}\bigg)\,,
}
and
\eq{
\tilde{\mathcal{Z}}_{1} = -{1\over \alpha}\left({1\over 2v}+\ln(v/2)+\gamma-2\sum^{\infty}_{p=1}K_0(2\pi p v)\right)\,.
}
Owing to the presence of the Bessel functions, the evaluation of the sums can be carried out numerically very easily.

When the value of $v^2$ is small, then we may proceed by splitting the summation range in two parts: one up to a value $\bar n \gg v^2$ plus a reminder. We can then expand the reminder for small $v^2$ and complete the infinite sums by adding and subtracting appropriate terms. The procedure is identical to that developed in Ref.~\cite{flachi} where the reader is addressed to check the details. The other regular term, involving  spurious divergent summations over $n$ is $\Delta_1$ that can also be treated along the same lines described above, i.e. expressing it in terms of regularized zeta functions.

The numerical procedure can be implemented straightforwardly and essentially it comes down to calculating every term of (\ref{phiterms}). The contributions $\Delta_1$, $\Delta_{2}$, $\Delta_{3}$, $\Delta_{4}$ and $\Upsilon_0$ do not pose any particular complication and their evaluation can be carried out directly. The term $\Sigma_1$ describes basically the reminder of the WKB approximation and it can be calculated by first solving the homogeneous equation numerically, followed by the subtraction of the approximate result using the WKB expansion up to leading order. The WKB expansion improves for large $l$ and $n$, however, computationally, this term is the most expensive to calculate. For $\Sigma_1$ we expedited this procedure by using a sampling method to compute the sums. Finally, the term $\Sigma_2$ consists of the sum of $\mathcal{P}_1$, $\mathcal{P}_2$ and $\mathcal{P}_3$.  Since we have explicitly shown its regularity, we may substitute the diverging functions $\mathcal{Z}_{q}$ (for $q=-1, 1$) with their regularized counterparts (\ref{regz}), while the functions $\mathcal{Z}_{q}$ (for $q>1$) can be evaluated easily due to the fast convergence of the sums over the Bessel functions. With these preliminaries, the terms $\mathcal{P}_1$ and $\mathcal{P}_2$ can, then, be calculated directly. The remaining term $\mathcal{P}_3$ can be calculated by adding and subtracting the expansion (\ref{expP3}) to (\ref{P3}) and using the WKB approximation for the terms containing $W_n$, which are then expanded in a Taylor series up to some order $j_{max}$. The advantage of using this approach is that the subtraction term quickly drops to zero for relatively small value of $j_{max}$, leading to a faster numerical convergence.

So far the treatment has been independent of the explicit form of the metric functions, so now we must specify them. Some examples are illustrated in Figs.~1 and 2, where we considered
\eq{
f = {1 \over u} = 1- \left({r_h \over r}\right)^{z+3}
}
for which the black hole temperature is given by
\eq{
T_{\textrm{bh}} = {1 \over 4\pi} {g_{00,1} \over \sqrt{g_{11} \, g_{00}}}\bigg|_{r=r_h} = {z+3 \over 4\pi} \, r_h^z\,.
}
The plots for the vacuum polarization are for the specific values $\xi=0$, $m=0.01$, with $z=1$ and $z=2$, for different values of horizon radius $r_0$.

\begin{figure}
\resizebox{0.4\textwidth}{!}{
\begin{tikzpicture}
  \begin{axis}[
    xlabel=$r-r_h$,
    ylabel={$\braket{\phi^2}$},
    xmin=0,
    xmax=100,
    ymin=0.12,
    ymax=0.25,
    legend style = {draw=none}
  ] 

\addplot[mark=none] coordinates{
(0.067672,0.117465)
(0.153662,0.146014)
(0.262928,0.168431)
(0.401772,0.184771)
(0.578199,0.195711)
(0.802382,0.202153)
(1.08725,0.20504)
(1.44923,0.205241)
(1.90919,0.203489)
(2.49365,0.20036)
(3.23633,0.196282)
(4.18003,0.191567)
(5.37919,0.186432)
(6.90294,0.181031)
(8.83916,0.175472)
(11.2995,0.169843)
(14.4258,0.164218)
(18.3984,0.158683)
(23.4462,0.153344)
(29.8605,0.148341)
(38.0111,0.143854)
(48.3679,0.140083)
(61.5282,0.137239)
(78.2508,0.13551)
(99.5,0.135015)
};

\addplot[mark=none, dashed] coordinates{
(0.0887776,0.184403)
(0.200074,0.193594)
(0.3396,0.201641)
(0.514517,0.207426)
(0.733802,0.210871)
(1.00871,0.212207)
(1.35335,0.211776)
(1.7854,0.209938)
(2.32704,0.207025)
(3.00608,0.203314)
(3.85734,0.199027)
(4.92454,0.194331)
(6.26243,0.189353)
(7.93967,0.184184)
(10.0423,0.178894)
(12.6784,0.173539)
(15.983,0.168169)
(20.1259,0.162841)
(25.3196,0.157621)
(31.8307,0.152599)
(39.9933,0.147884)
(50.2264,0.143612)
(63.0552,0.139942)
(79.1379,0.137063)
(99.3,0.135163)
};

\addplot[mark=none, dotted] coordinates{
(0.108477,0.211925)
(0.243105,0.213297)
(0.410185,0.215499)
(0.617542,0.216957)
(0.874885,0.217304)
(1.19426,0.216517)
(1.59063,0.214708)
(2.08255,0.212039)
(2.69305,0.208678)
(3.45071,0.204783)
(4.39102,0.200484)
(5.55801,0.195887)
(7.0063,0.191074)
(8.80373,0.186108)
(11.0344,0.18104)
(13.8029,0.17591)
(17.2387,0.170758)
(21.5028,0.165624)
(26.7947,0.160558)
(33.3624,0.155622)
(41.5132,0.15089)
(51.6289,0.146451)
(64.1831,0.142424)
(79.7636,0.138963)
(99.1,0.136249)
};

\addplot[mark=none, densely dotted] coordinates{
(0.127099,0.225481)
(0.28357,0.222998)
(0.476199,0.222302)
(0.713343,0.221598)
(1.00529,0.22039)
(1.3647,0.218539)
(1.80717,0.216047)
(2.35188,0.21298)
(3.02248,0.20943)
(3.84804,0.205489)
(4.86439,0.201241)
(6.11559,0.196757)
(7.65594,0.192097)
(9.55225,0.187307)
(11.8868,0.182426)
(14.7608,0.177487)
(18.299,0.17252)
(22.6548,0.16756)
(28.0171,0.162645)
(34.6187,0.15782)
(42.7458,0.153139)
(52.751,0.148665)
(65.0684,0.144485)
(80.2321,0.140722)
(98.9,0.137532)
};

\legend{$r_h = 0.5$, $r_h = 0.7$, $r_h = 0.9$, $r_h = 1.1$}

\end{axis}
\end{tikzpicture}
}
\caption{Vacuum polarization for $\xi = 0$, $m=0.01$ and $z=1$.}
\end{figure}
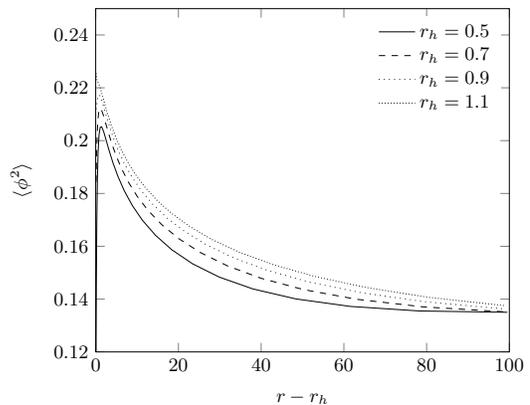

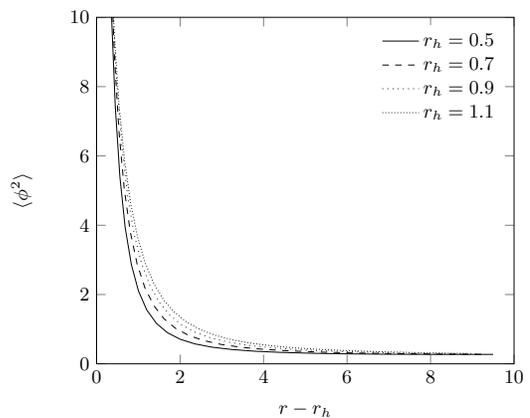
\begin{figure}
\resizebox{0.4\textwidth}{!}{
\begin{tikzpicture}
  \begin{axis}[
    xlabel=$r-r_h$,
    ylabel={$\braket{\phi^2}$},
    xmin=0,
    xmax=10,
    ymin=0,
    ymax=10,
    legend style = {draw=none}
  ] 

\addplot[mark=none] coordinates{
(0.039456,67.2502)
(0.0851392,40.0996)
(0.138032,27.2638)
(0.199273,19.3737)
(0.270179,14.007)
(0.352276,10.1882)
(0.44733,7.4184)
(0.557385,5.39864)
(0.68481,3.92875)
(0.832345,2.86526)
(1.00317,2.10152)
(1.20095,1.55719)
(1.42994,1.1718)
(1.69507,0.900266)
(2.00205,0.709387)
(2.35748,0.575086)
(2.76901,0.480175)
(3.24548,0.412605)
(3.79715,0.364097)
(4.43589,0.329131)
(5.17543,0.304201)
(6.0317,0.287239)
(7.0231,0.277008)
(8.17097,0.272512)
(9.5,0.272687)
};

\addplot[mark=none, dashed] coordinates{
(0.0496561,51.8947)
(0.106357,31.4775)
(0.171102,21.8648)
(0.245034,15.9304)
(0.329454,11.8488)
(0.425851,8.89359)
(0.535924,6.70022)
(0.661614,5.05504)
(0.805137,3.81805)
(0.969021,2.89001)
(1.15616,2.19698)
(1.36984,1.68236)
(1.61384,1.30233)
(1.89246,1.02303)
(2.21061,0.818411)
(2.5739,0.668678)
(2.98872,0.558944)
(3.4624,0.47817)
(4.00328,0.418289)
(4.6209,0.373515)
(5.32615,0.339787)
(6.13145,0.314353)
(7.051,0.29546)
(8.10102,0.282083)
(9.3,0.273613)
};

\addplot[mark=none, dotted] coordinates{
(0.058492,43.0589)
(0.124587,26.4305)
(0.199273,18.6421)
(0.283667,13.8271)
(0.379031,10.4942)
(0.48679,8.05457)
(0.608556,6.21647)
(0.74615,4.81157)
(0.901628,3.73155)
(1.07732,2.90071)
(1.27584,2.26302)
(1.50017,1.77545)
(1.75366,1.4043)
(2.04009,1.12293)
(2.36376,0.910356)
(2.7295,0.750074)
(3.14278,0.629249)
(3.60977,0.537989)
(4.13747,0.468755)
(4.73376,0.41588)
(5.40756,0.375171)
(6.16894,0.343589)
(7.02929,0.318994)
(8.00146,0.299947)
(9.1,0.28556)
};

\addplot[mark=none, densely dotted] coordinates{
(0.0662621,37.3026)
(0.140507,23.0987)
(0.223697,16.4838)
(0.316909,12.395)
(0.421352,9.55386)
(0.538377,7.45892)
(0.669501,5.86375)
(0.816422,4.62793)
(0.981044,3.66235)
(1.1655,2.90561)
(1.37218,2.31272)
(1.60375,1.84922)
(1.86323,1.48804)
(2.15397,1.20756)
(2.47973,0.990419)
(2.84474,0.822711)
(3.25373,0.693324)
(3.71199,0.593459)
(4.22546,0.516198)
(4.80079,0.456165)
(5.44544,0.409226)
(6.16774,0.372252)
(6.97708,0.342921)
(7.88391,0.319552)
(8.9,0.30098)
};

\legend{$r_h = 0.5$, $r_h = 0.7$, $r_h = 0.9$, $r_h = 1.1$}
\end{axis}
\end{tikzpicture}
}
\caption{Vacuum polarization for $\xi = 0$, $m=0.01$ and $z=2$.}
\end{figure}

As a simple consistency check, we may investigate the asymptotic values for large values of the radial coordinate in the case $z=1$. In such a limit, it is readily seen that the metric functions reduce to an AdS type, for which we know that the analytic asymptotic value of the vacuum polarization is given by
\eq{
\braket{\phi^2}_{\textrm{AdS}} \approx -{1 \over 48 \pi^2}\,.
}
We verify that this is indeed the value obtained in our case as well.

\section{Conclusions}

Lifshitz black holes are black hole solutions exhibiting scaling. These solutions are important ingredients in the construction of gravitational dual of Lifshitz field theories, allowing one to investigate finite temperature effects. In this paper we have addressed the problem of calculating the coincidence limit of the Green function for a massive, non-minimally coupled bulk scalar field, i.e., the vacuum polarization $\langle \phi^2 \rangle$. The computation of quantum vacuum effects is a notoriously difficult task and, here, we have adapted the methodology used in the case of asymptotically anti-de Sitter black holes to the case of Lifshitz black holes. 

The basic approach relies on the use of the WKB approximation and point splitting regularization together, allowing us to express the full solution as a WKB approximated part plus a remainder. This proves to be very effective to explicitly confirm the cancellation of diverging parts, while at the same time providing a regular set up for a numerical calculation. The WKB part is directly computed by using the analytic results expressed in terms of regulated generalized zeta functions, which in turn converge very rapidly due to the fast decay of the modified Bessel functions appearing in them. The remainder part is calculated by numerically solving the mode equation and subtracting the WKB counterparts. The convergence in this case is quite fast as well, since this component is of order $\mathcal{O}(l^{-5},n^{-5})$.

We have dealt with the most general case of spherically symmetric Lifshitz solutions. We then have considered a particular form for the metric functions in order to obtain a numerical result. We chose a function which asymptotes to an AdS case, for which an analytic result had already been calculated, and used it to check with our results which correctly reproduced the expected behavior.

\acknowledgments
We thank Funda\c c\~ao para a Ci\^encia e Tecnologia (FCT) - Portugal
for financial support through Project No.~PEst-OE/FIS/UI0099/2015 and
the European Union Seventh Framework Programme through Grant Agreement
PCOFUND-GA-2009-246542.  GQ also acknowledges the grant
No. SFRH/BD/92583/2013 from FCT. 
AF kindly acknowledges the MEXT-Supported Program for the Strategic Research Foundation at Private Universities `Topological Science' (Grant No. S1511006).

\end{document}